\documentclass[aps,twocolumn,showpacs]{revtex4-1}
\usepackage{bm}
\usepackage{color} 
\usepackage{graphicx}
\usepackage{epsfig}
\usepackage{hyperref}
\usepackage{natbib}
\usepackage{amsmath}
\usepackage{amssymb}

\newcommand{\av}[1]{\left< {#1} \right>}

\begin{document}

\title{Effect of Dresselhaus spin-orbit coupling on spin dephasing in asymmetric and macroscopically symmetric (110)-grown quantum wells} 

\author{A.\,V.\,Poshakinskiy}
\author{S.\,A.\,Tarasenko}

\affiliation{A.\,F.\,Ioffe Physical-Technical Institute, Russian Academy of Sciences, 194021 St.~Petersburg, Russia}


%
%
%
%
%
%
%
%

\begin{abstract}
We develop the microscopic theory of electron spin dephasing in (110)-grown quantum wells
where the electron scattering time is comparable to or exceeds the period of spin precession in the effective magnetic field caused by spin-orbit coupling. Structures with homogeneous and fluctuating Rashba field, which triggers the dephasing of electron spins aligned along the growth direction, are analyzed. We show that the Dresselhaus field, which is always present in zinc-blende-type quantum wells, suppresses the spin dephasing enabling very long spin lifetime of conduction electrons. The dependence of the spin lifetime on the electron mobility is found to be nonmonotonic reaching the minimum in structures where the scattering time is comparable to the period of spin precession in the effective magnetic field.  
\end{abstract}

\maketitle

\section{Introduction}

The spin dynamics of free electrons in zinc-blende-type quantum wells (QWs) is largely determined by the electron subband spin-orbit splitting which can be fruitfully treated as a wavevector-dependent effective magnetic field acting upon the spins of individual electrons~\cite{Dyakonov86,Gridnev01,Averkiev02,Wu10}. The field occurs due to the lack of space inversion symmetry in a particular QW and can originate from bulk, structure, or interface inversion asymmetry, or their joint action. The field direction and magnitude depend on the QW crystallographic orientation and design as well as external gate voltage~\cite{Ohno99,Karimov03,Belkov08,Olbrich09,Eldridge11,English11,Balocchi11,Ye12}, which opens a way towards the spin manipulation and spin-based semiconductor devices. Of special interest are QWs grown on the (110) substrate. The effective magnetic field caused by bulk inversion asymmetry (Dresselhaus field) in such structures points along the growth direction and, therefore, does not cause the precession of electron spins oriented along the QW normal~\cite{Dyakonov86}. This enables the observation of very long electron spin lifetimes up to hundreds of nanoseconds at low temperatures which are unfeasible for two-dimensional structures of other crystallographic orientations~\cite{Dohrmann04,Couto07,Muller08,Volkl11,Griesbeck12}. 

The electron spin lifetime in (110)-grown QWs at low temperature is likely to be limited by the spin precession in the 
homogeneous Rashba field present in asymmetric QWs or unavoidable fluctuating Rashba field that emerges due to domain structure formation or inhomogeneous distribution of charged impurities in doped structures~\cite{Sherman03}. The Rashba field lies in the QW plane and starts up the spin dephasing. The previous theoretical studies of electron spin dephasing in (110)-grown QWs with the homogeneous~\cite{Cartoixa05,Tarasenko09} or fluctuating~\cite{Glazov_PRB10,Zhou10,Glazov10} Rashba field were limited to the collision-dominated regime of spin dynamics, where the scattering time is much smaller than the spin precession period in the effective magnetic field. Such a consideration is inapplicable for QWs with high electron mobility and/or strong spin-orbit coupling, which are now technological available.
In this paper, we develop the microscopic theory of electron spin dephasing for arbitrary ratio between the scattering time and the spin precession period. We show that the interplay of Rashba and Dresselhaus fields leads to dynamic coupling of the in-plane and out-of-plane spin components. The dependence of the spin lifetime on the electron mobility is nonmonotonic and reaches the minimum in QW structures where the scattering time is comparable to the period of spin precession in the effective magnetic field. In high-mobility QWs, the Dresselhaus field efficiently suppresses the spin dephasing allowing one to reach very long spin lifetimes.      

\section{Uniform Rashba field}

First, we consider asymmetric (110)-grown QWs with a spatially homogeneous Rashba field. The linear in the wavevector  spin-orbit splitting of the electron subband in such QWs ($C_s$ point group) is generally described by three linearly independent parameters $\alpha_1$, $\alpha_2$, and $\beta$~\cite{Nestoklon12} (see also Refs.~\cite{Cartoixa06,Shalygin06}). The corresponding Hamiltonian has the form
\begin{equation}\label{ham}
H_{so} =  \alpha_1 \sigma_x k_y  -\alpha_2 \sigma_y k_x + \beta \sigma_z k_x \:,
\end{equation}
where $\sigma_j$ ($j=x,y,z$) are the Pauli matrices, $k_j$ are the components of the electron wavevector, $x \parallel [1\bar 10]$ and $y \parallel [00\bar 1]$ are the in-plane axes, and $z \parallel [110]$ is the QW normal. The parameters $\alpha_1$ and $\alpha_2$ are nonzero in QWs with structure inversion asymmetry only, while $\beta$ requires bulk inversion asymmetry and describes the Dresselhaus splitting in (110)-oriented QWs. The difference between $\alpha_1$ and $\alpha_2$ is usually small because it is caused by an interference effect between cubic structure of the crystall lattice and QW
structure inversion asymmetry~\cite{Nestoklon12}. Therefore, we assume in what follows that $\alpha_1=\alpha_2=\alpha$. The Hamiltonian~(\ref{ham}) is similar to the Zeeman term with the effective Larmor frequency
\begin{equation}\label{omega110}
\bm{\Omega_k} = \left(  \Omega_R\frac{k_y}{k_F}, -\Omega_R\frac{k_x}{k_F}, \Omega_D\frac{k_x}{k_F} \right) \:,
\end{equation}
where $\Omega_R = 2 \alpha k_F /\hbar$ and $\Omega_D=2 \beta k_F /\hbar$ are the frequencies corresponding to Rashba and maximal Dresselhaus splitting at the Fermi level, and $k_F$ is the Fermi wavevector. Following the experiments~\cite{Belkov08,Olbrich09,Muller08,Griesbeck12} we consider the degenerate electron gas and study the spin dynamics at the Fermi level.
\begin{figure}
\includegraphics[width=0.95\columnwidth]{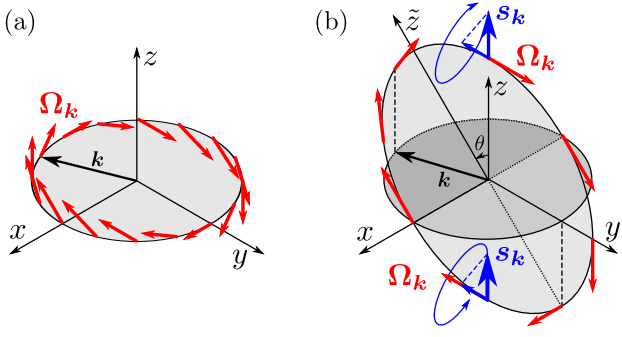}
\caption{(a) Dependence of the Larmor frequency $\bm{\Omega_k}$ corresponding to the effective magnetic field on the wavevector $\bm{k}$ at the Fermi circle. (b) The $(x \tilde z)$ plane containing $\bm{\Omega_k}$ for all the wavevectors $\bm k$. Precession of electron spins initially aligned along the QW normal $z$ with the frequency $\bm{\Omega_k}$
leads to the emergence of an average spin component along $y$.}
\label{fig1}
\end{figure}

Figure~\ref{fig1}a shows the distribution of the frequency $\bm{\Omega_k}$ on the wavevector at the Fermi level. The distribution is strongly anisotropic; the frequency $\bm{\Omega_k}$ contains both the in-plane and out-of-plane components. However, one can notice that the frequencies $\bm{\Omega_k}$ corresponding to various wavevectors all lie in a certain plane $(x \tilde z)$, see Fig~\ref{fig1}b. The $(x \tilde z)$ plane is obtained from $(xz)$ by the frame rotation around the $x$ axis with the angle $\theta=\arctan (\Omega_R/\Omega_D)$. 

\subsection{Time dependence}

The fact that the plane of the $\bm{\Omega_k}$ vectors does not contain the growth direction $z$ nor is perpendicular to $z$ is a feature of low-symmetry QWs and leads to the coupling of the in-plane and out-of-plane spin components. This is illustrated in Fig.~\ref{fig1}b which shows the precession of electron spins initially oriented along the QW normal. In the ballistic regime, i.e., the absence of scattering, the time dependence of electron spin at each point $\bm{k}$ on the Fermi circle is
given by
\[
\bm{s_k}(t) = \frac{\bm{\Omega_k} \cdot \bm{s}(0)}{\Omega_{\bm{k}}^2} \bm{\Omega_k} + \left[ \bm{s}(0) - \frac{\bm{\Omega_k} \cdot \bm{s}(0)}{\Omega_{\bm{k}}^2} \bm{\Omega_k} \right] \cos (\Omega_{\bm{k}} t)
\]
\vspace{-3mm}
\begin{equation}\label{s_k_t}
+ \, \frac{\bm{\Omega_k} \times \bm{s}(0)}{\Omega_{\bm{k}}} \sin (\Omega_{\bm{k}} t) \:,
\end{equation}
where $\bm{s}(0)$ is the initial spin. The time evolution of the components of the total electron spin
\begin{equation}
\bm{S}(t) = \sum_{\bm{k}} \bm{s_k} (t) 
\end{equation}
is obtained by summation $\bm{s_k} (t)$ over the wavevectors at the Fermi level, which yields
\[
S_j (t) = \sum_{j'} \av{\frac{\Omega_{\bm{k}, j} \, \Omega_{\bm{k},j'}}{\Omega_{\bm{k}}^2}} S_{j'}(0)  
\]
\vspace{-3mm}
\begin{equation}\label{noscat}
+ \sum_{j'}
\av{\frac{\Omega_{\bm{k}}^2 \, \delta_{jj'} - \Omega_{\bm{k}, j} \, \Omega_{\bm{k},j'}}{\Omega_{\bm{k}}^2} \cos (\Omega_{\bm{k}} t) } S_{j'}(0) \:,
\end{equation}
where the angle brackets denote averaging over the direction of $\bm{k}$. The first term on the right-hand side of Eq.~(\ref{noscat}) presents the averaged projection of individual electron spins onto the effective field direction and is time-independent. In (110) QWs, this contribution is present even for the spin polarized along the QW normal, which is in contrast to (001)-grown structures where the effective field lies in QW plane and the time-independent term vanishes for $\bm{S}(0) \parallel z$. The second term on the right-hand side of Eq.~(\ref{noscat}) describes spin oscillations.

\begin{figure}[t]
\includegraphics[width=0.95\columnwidth]{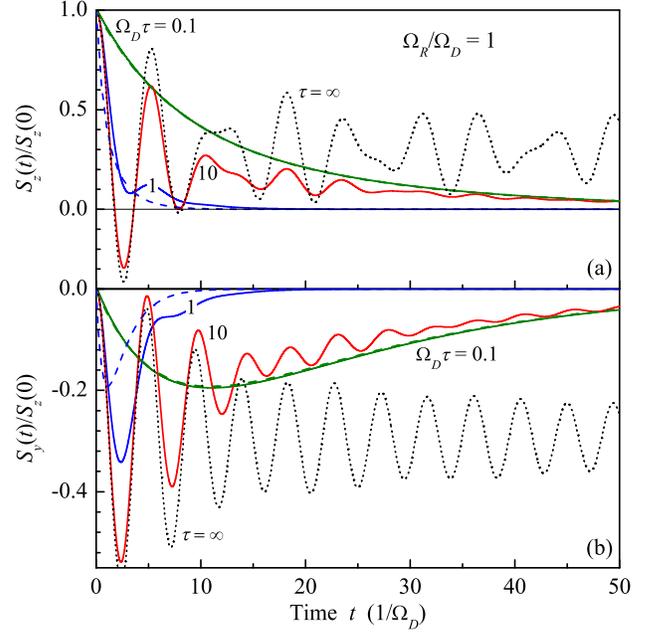}
\caption{Time dependence of the components of the total electron spin $\bm{S}(t)$ oriented along the QW normal $z$ at $t=0$. Curves are plotted for $\Omega_R=\Omega_D$ and different parameters $\Omega_D \tau$ characterizing the electron mobility. Dotted curves correspond to ballistic transport, $\tau = \infty$, and are plotted after Eq.~(\ref{noscat}). 
Dashed curves present the results of the collision-dominated approximation, $\Omega_D \tau \ll 1$, and are calculated after Eq.~(\ref{eq:coldomt}).}
\label{fig:time}
\end{figure}
Time dependence of the out-of-plane and in-plane components of the total spin oriented along the QW normal at $t=0$ is shown by dotted curves in Fig.~\ref{fig:time}a and~\ref{fig:time}b. One can see that, in asymmetrical (110)-grown QWs, the $S_z$ and $S_y$ projections are dynamically coupled: the dephasing of spins polarized along the QW normal leads to the emergence of the in-plane spin component. Both $S_z(t)$ and $S_y(t)$ contain oscillations which slowly decay in time. Analytical expressions for  $S_z(t)$ and $S_y(t)$ at $t > 1/\Omega_{\bm{k}}$ can be derived by the stationary phase method, which gives
\begin{eqnarray}
\frac{S_z(t)}{S_z(0)} \approx \left( 1-\frac{\Omega_R}{\Omega_\Sigma} \right) + \sqrt{\frac{2 \Omega_R}{\pi \Omega_D^2 t}} \, \cos \left( \Omega_R t +\frac{\pi}{4} \right) \,\\
+ \sqrt{\frac{2 \Omega_R^4}{\pi \Omega_D^2 \Omega_\Sigma^3 t}} \, \cos \left( \Omega_\Sigma t -\frac{\pi}{4} \right) \nonumber ,
\end{eqnarray} 
\[
\frac{S_y(t)}{S_z(0)} \approx - \left( 1-\frac{\Omega_R}{\Omega_\Sigma} \right) \frac{\Omega_R}{\Omega_D} + \sqrt{\frac{2 \Omega_R^2}{\pi \Omega_\Sigma^3 t}} \, \cos \left( \Omega_\Sigma t -\frac{\pi}{4} \right) ,
\]
where $\Omega_\Sigma = \sqrt{\Omega_R^2 + \Omega_D^2}$. The oscillation frequencies are determined by the smallest ($\Omega_R$) and the largest ($\Omega_{\Sigma}$) magnitudes of the effective Larmor frequency at the Fermi circle.
At $t \rightarrow \infty$, the oscillations vanish and the spin polarization reaches a stationary value. 

The scattering of electrons by static defects or phonons in real structures leads to the isotropization of spin distribution in $\bm{k}$-space which drastically suppresses the spin oscillations. In the presence of scattering, the spin dynamics can be described by the kinetic equation 
\begin{equation}\label{eq:kin}
\frac{\partial \bm{s_k}}{\partial t} + \bm{s_k} \times \bm{\Omega_k} = {\bm g} -\frac{\bm{s_k} - \langle \bm{s_k} \rangle}{\tau} \:,
\end{equation}
where $\bm{g}$ is the spin generation rate at the Fermi level, e.g., due to resonant optical pumping with circularly polarized light, and $\tau$ is the scattering time. We assume that $\bm{g}$ is independent of the wavevector direction. The last term on the right-hand side of Eq.~(\ref{eq:kin}) presents a simple form of the collision integral. Effects of scattering anisotropy on spin dephasing were studied in Ref.~\cite{Poshakinskiy11a}. 

Solid curves in Figs.~\ref{fig:time}a and~\ref{fig:time}b show the time evolution of the out-of-plane and in-plane projections of the total electron spin $\bm{S}(t)$ oriented at $t=0$ along the QW normal. The curves are calculated numerically after Eq.~(\ref{eq:kin}) for different scattering times $\tau$. As expected, the decrease in the scattering time results in a suppression of spin oscillations. Surprisingly, the effect of electron scattering on spin lifetime is more complicated. Comparison of different curves in Fig.~\ref{fig:time}a and~\ref{fig:time}b shows that the shortest spin lifetime $T \sim \tau$ is achieved at a certain mobility corresponding to $\tau \sim 1/\Omega_{\bm{k}}$. At both higher, $\tau >  1/\Omega_{\bm{k}}$, and lower, $\tau < 1/\Omega_{\bm{k}}$, mobility the spin lifetime increases as $T \propto \tau$ and $T \propto 1/(\Omega_{\bm{k}}^2 \tau)$, respectively. 

In the collision-dominated regime, $\Omega_R \tau,\, \Omega_D\tau \ll 1$, one can analytically solve Eq.~(\ref{eq:kin})
and obtain the master equation for the total electron spin~\cite{Dyakonov86}
\begin{equation}\label{kintot}
\frac{d\bm{S}}{dt} = \bm{G} - \bm{\Gamma}_{\rm cd} \bm{S} \:,
\end{equation}
where $\bm{G}=\sum_{\bm{k}} \bm{g}$ and $\bm{\Gamma}_{\rm cd}$ is the D'yakonov-Perel' spin-relaxation-rate tensor,
$(\Gamma_{{\rm cd}})_{jj'} = \tau \av{\Omega_{\bm{k}}^2 \delta_{jj'} - \Omega_{\bm{k},j} \, \Omega_{\bm{k},j'}}$. For (110)-grown QWs, the tensor $\bm{\Gamma}_{\rm cd}$ has the form~\cite{Tarasenko09}
\begin{eqnarray}\label{gammacd}
\bm{\Gamma}_{\rm cd} &=& \frac{\tau}{2} \left[ \begin{array}{ccc} {\Omega_R^2+\Omega_D^2} & 0 & 0 \\  %
                                                       0 & {\Omega_R^2+\Omega_D^2} & {\Omega_R \Omega_D} \\  %
                                                       0 &  {\Omega_R \Omega_D} & 2 \Omega_R^2  \end{array} \right] . \end{eqnarray}
Solution of Eq.~(\ref{kintot}) with $\bm{G}=0$ and the initial condition $\bm{S}(0) \parallel z$ has the biexponential
form
\begin{eqnarray}\label{eq:coldomt}
S_z (t)  &=&   \frac{\Omega_R^2 \exp( - \gamma_{\tilde{y}} t)  + \Omega_D^2 \exp(-\gamma_{\tilde{z}} t) }{\Omega_{R}^2 + \Omega_{D}^2}  S_z(0) \:, \nonumber \\
S_y (t) &=&  \frac{\Omega_R \Omega_D \left[ \exp( - \gamma_{\tilde{y}} t) - \exp(-\gamma_{\tilde{z}} t) \right]}{\Omega_{R}^2 + \Omega_{D}^2} \, S_z(0) \: ,
\end{eqnarray}
where $\gamma_{\tilde{y}} = (2\Omega_R^2 + \Omega_D^2)\tau/2$ and $\gamma_{\tilde{z}} = \Omega_R^2 \tau/2$ are eigen values of the $\bm{\Gamma}$ tensor. The functions~(\ref{eq:coldomt}) are plotted in Figs.~\ref{fig:time}a and~\ref{fig:time}b by dashed curves and demonstrate the perfect agreement with the results of numerical calculation for the collision-dominated regime.

\subsection{Continuous-wave excitation}

Besides experiments with time resolution, spin dephasing is also widely studied at continuous-wave (cw) pumping. In the regime linear in the pump intensity, the general relation between the steady-state spin $\bm S$ and the spin generation rate $\bm G$ has the form
\begin{equation}\label{eq:Tdef}
\bm S = \bm T \bm G \:,
\end{equation}
where $\bm{T}$ can be referred to as the spin lifetime tensor. To calculate 
$\bm{T}$ 
we solve Eq.~(\ref{eq:kin}) for the stationary case following the procedure described in Ref.~\cite{Tarasenko06}. We express the spin distribution function $\bm{s}_{\bm{k}}$ via its average value
\begin{equation}\label{eq:sol1}
{\bm{ s_k}} = \frac{ \bm{\zeta}_{\bm k} +  \tau \, \bm{\Omega_k} \times \bm{\zeta}_{\bm k} + \tau^2 \, \bm{\Omega_k} (\bm{\Omega_k} \cdot \bm{\zeta}_{\bm k})}{1 + \bm{\Omega_k}^2 \tau^2} \:,
\end{equation}
where $\bm{\zeta}_{\bm k} = \av{\bm{s_k}} + {\bm{g}}\tau$, sum up Eq.~(\ref{eq:sol1}) over the wavevector and obtain the linear equation set
\begin{equation}\label{eq:sol2}
 \sum_{j'} \av{\frac{\Omega_{\bm{k}}^2 \, \tau \, \delta_{jj'} - \Omega_{\bm{k},j} \, \Omega_{\bm{k},j'} \, \tau}{1 + \Omega_{\bm{k}}^2 \tau^2}} (S_{j'} + G_{j'} \tau )  = G_j \:.
\end{equation}
Solution of the equation set~(\ref{eq:sol2}) for the effective Larmor frequency~(\ref{omega110}) yields
\begin{equation}\label{eq:T}
\bm{T} = \dfrac{\tau}{\sqrt{1+\Omega_R^2 \tau^2}\sqrt{1+\Omega_\Sigma^2 \tau^2}-1} 
\end{equation}
\[
\times \left[ 
\begin{array}{ccc} 1 +\dfrac{\Omega_R^2(1+\Omega_{\Sigma}^2\tau^2)}{\Omega_{\Sigma}^2} \hspace{-10mm} & \hspace{10mm} 0 & 0 \\ 
0 & 2+\Omega_R^2 \tau^2 & - \dfrac{\Omega_D (1+\Omega_R^2\tau^2)}{\Omega_R } \\  %
0 & - \dfrac{\Omega_D (1+\Omega_R^2 \tau^2)}{\Omega_R } & \Omega_D^2\tau^2 + \dfrac{\Omega_{\Sigma}^2}{\Omega_R^2}
\end{array} \right] .
\] 
Equation~(\ref{eq:T}) is one of the principal results of the paper. It enables one to calculate the steady-state spin polarization for arbitrary ratio between the scattering time $\tau$ and the precession periods $1/\Omega_R$, $1/\Omega_D$ and shows that the Dresselhaus field considerably affects the electron spin dephasing in (110)-grown QWs.  

\begin{figure}
\includegraphics[width=0.95\columnwidth]{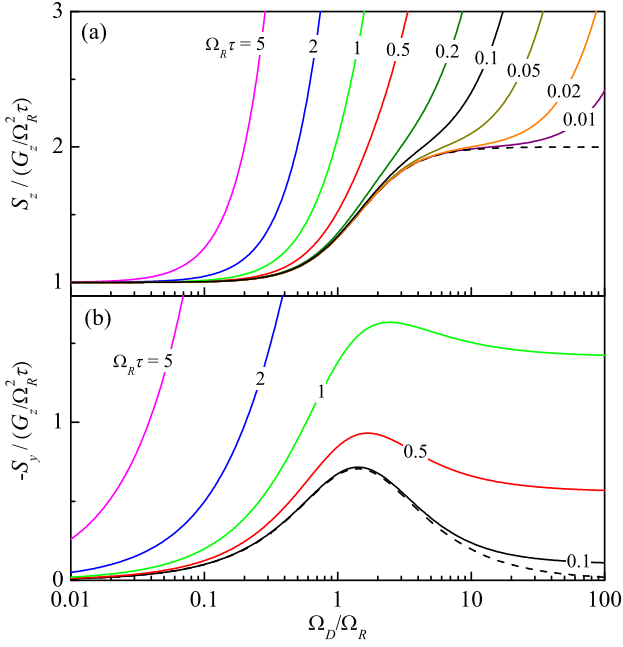}
\caption{Dependence of the steady-state spin components $S_z$ and $S_y$ on the Dresselhaus field for cw spin pumping along the QW normal $z$. Solid curves are plotted for different $\Omega_R \tau$ and normalized by $G_z/(\Omega_R^2\tau)$ which determines $S_z$ at $\Omega_D=0$. Dashed curves correspond to the collision-dominated regime, $\Omega_R,\Omega_D \ll 1/\tau$.}
\label{fig:dress}
\end{figure}

Figures~\ref{fig:dress}a and~\ref{fig:dress}b show the dependence of the $S_z$ and 
$S_y$ projections of the steady-state spin, caused by excitation along the QW normal $z$, on the Dresselhaus field. The curves are calculated after Eqs.~(\ref{eq:Tdef}) and~(\ref{eq:T}) for different $\Omega_R\tau$. Despite the fact that the Dresselhaus field points along the QW normal, it efficiently suppresses the spin dephasing increasing $S_z$ as well as leads to the emergence of $S_y$. The $S_z$ projection (Fig.~\ref{fig:dress}a) is determined by the $T_{zz}$ component of the spin lifetime tensor. The latter equals to $1/(\Omega_R^2 \tau)$ at $\Omega_D=0$ and increases with the Dresselhaus field as $(\Omega_D/\Omega_R)\sqrt{1+\Omega_R^2 \tau^2}$ at $\Omega_D \tau \gg 1$. Such an increase in the spin lifetime is caused by stabilizing action of the Dresselhaus field on the normal spin component. The in-plane spin projection (Fig.~\ref{fig:dress}b) is determined by the off-diagonal component $T_{yz}$ which nonmonotonously depends on the Dresselhaus field. This component linearly increases with the Dresselhaus field as ${-\Omega_D(1+\Omega_R^2\tau^2)/(\Omega_R^3\tau)}$ at small $\Omega_D$, reaches a peak value of ${-\tau(1+\Omega_R^2\tau^2)^{3/2}/[\Omega_R^2\tau^2 \sqrt{2+\Omega_R^2\tau^2}]}$ at ${\Omega_D^*=\Omega_R \sqrt{(1+\Omega_R^2\tau^2)(2+\Omega_R^2\tau^2)}}$, and then saturates to $-(1+\Omega_R^2\tau^2)/\Omega_R$ at $\Omega_D\tau \gg 1$.

\begin{figure}
\includegraphics[width=0.95\columnwidth]{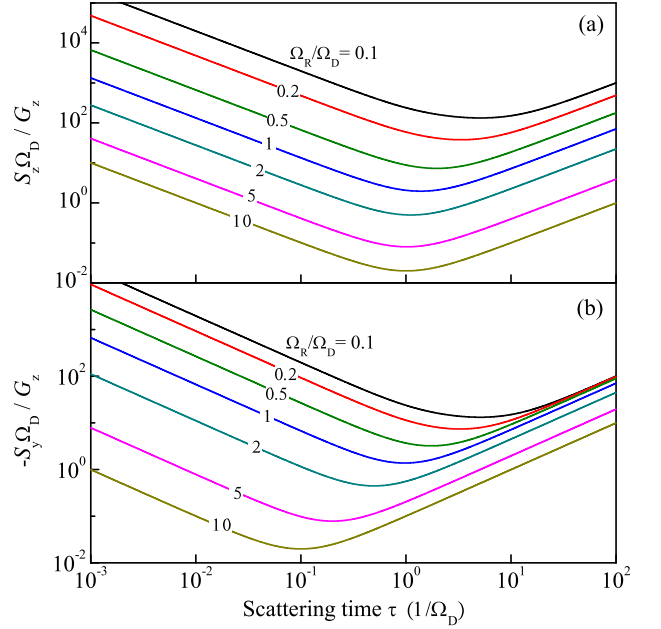}
\caption{Dependence of the steady-state spin components $S_z$ and $S_y$ on the scattering time $\tau$ for cw spin pumping along the QW normal $z$. Curves correspond to different ratio $\Omega_R/\Omega_D$.}
\label{fig:mob}
\end{figure}

The dependence of the steady-state projections $S_z$ and $S_y$ on the scattering time $\tau$ is shown Fig.~\ref{fig:mob}.
In low mobility structures, the spin polarization decreases with the increase in $\tau$, in accordance with the theory of D'yakonov-Perel' spin relaxation in the collision-dominated regime. In particular, the spin lifetime tensor components determining $S_z$ and $S_y$ in this regime are given by $T_{zz}\approx(\Omega_R^2+\Omega_D^2)/[\Omega_R^2 (\Omega_R^2+\Omega_D^2/2) \tau]$ and $T_{yz}\approx-\Omega_D/[\Omega_R (\Omega_R^2+\Omega_D^2/2) \tau]$. At further increase in the scattering time, the spin polarization reaches a minimum and then rises linearly with $\tau$, see Fig.~\ref{fig:mob}. Such a behavior is caused by the transition from the collision-dominated to oscillatory regime of spin dynamics, see Fig.~\ref{fig:time}. In the oscillatory regime, the $T_{zz}$ and $T_{yz}$ components are given by $\Omega_D^2 \tau/(\Omega_R \sqrt{\Omega_R^2+\Omega_D^2})$ and 
$-\Omega_D\tau/\sqrt{\Omega_R^2+\Omega_D^2}$, respectively. We note that the nonmonotonic dependence $S_z(\tau)$ is a feature of low-symmetry QWs and absent, e.g., in (001) QWs where $S_z \propto 1/\tau$ in both collision-dominated and oscillatory regimes.

\section{Fluctuating Rashba field}

Now, we analyze the spin dephasing in macroscopically symmetric QWs where the Rashba field arises due to QW domain structure or inhomogeneous distribution of charged impurities in the doping layers~\cite{Sherman03,Grncharova76,Golub04} and fluctuates in the QW plane around its zero average value. In such systems, the Larmor frequency $\bm{\Omega_{k}}$ corresponding to the total effective field is given by the sum of $\bm\Omega_{D,\bm{k}}=\Omega_D(0,0,k_x/k_F)$ and position-dependent ${\bm{\Omega}_{R,\bm{k}}(\bm{\rho})=\Omega_{R}(\bm{\rho})(k_y/k_F,-k_x/k_F,0)}$ with $\bm\rho$ being the in-plane coordinate. The fluctuating Rashba field is usually weak, therefore, we assume that ${|\Omega_R(\bm{\rho})| \ll |\Omega_D|, 1/\tau}$. 

During the spin dephasing, electrons travel in the QW plane and their spins are affected by the Rashba field of different strength and sign. The characteristic time of the Rashba field change for an individual electron $\tau_c$ (correlation time) is determined by the domain size and electron mobility. Depending on the ratio between $\tau_c$ and other relevant times, different scenarios of spin dephasing are realized. Below, we consider them for the initial condition of spin polarization along the QW normal.

(i) Large domains, $\tau_c \gg 1/(\Omega_R^2 \tau) \gg 1/(\Omega_D^2 \tau), \, \tau$. Spin dephasing occurs in all domains
independently. To calculate the time evolution of electron spin in each domain, we solve kinetic Eq.~(\ref{eq:kin}) with $\bm{g}=0$ and the initial condition $\bm{s}_{\bm{k}}(0)\parallel z$. At $\Omega_R/\Omega_D \ll 1$, the $\tilde{z}$ axis is nearly parallel to the $z$ axis, see Fig.~\ref{fig1}(b). As the result, the in-plane component of the total electron spin in the domain is small. The out-of-plane component exhibits exponential decay 
\begin{equation}
S_z(t) = S_z(0) \exp(- \gamma_{\tilde z} t) \:, 
\end{equation}
with the dephasing time
\begin{equation}\label{eq:Gtildz}
\gamma_{\tilde{z}} = \frac{\Omega_R^2(\bm{\rho}) \tau}{1+\sqrt{1+\Omega_D^2\tau^2}} \:.
\end{equation}

The net electron spin averaged over all domains is given by
\begin{equation}
\overline{S_z(t)} =  \int\limits_{-\infty}^{\infty}  S_z(t) \, w(\Omega_R) \, d\Omega_R \:, 
\end{equation}
where $w(\Omega_R)$ is the distribution function of the Rashba field in the QW plane. Different dephasing rates $\gamma_{\tilde z}$ in domains
lead to a non-exponential decay of the net electron spin. In particular, at large times ($t \gg 1/\overline{\gamma_{\tilde{z}}}$ but still $t \ll \tau_c$), the time dependence of the net spin has the form
\begin{equation}
\overline{S_z(t)} = w(0) \left[ \frac{\pi(1+\sqrt{1+\Omega_D^2\tau^2})}{\tau t} \right]^{1/2} \hspace{-0.3cm} S_z(0) \:.
\end{equation}
The presence of the Dresselhaus spin-orbit coupling leads to a slowdown of spin dephasing in all domains and, hence, to an increase in the net spin polarization. 

(ii) $1/(\Omega_R^2 \tau) \gg \tau_c \gg 1/(\Omega_D^2 \tau), \, \tau$. The process of spin dephasing occurs in different domains at different rates $\gamma_{\tilde z}$, however, electrons frequently move from one domain to another. Therefore, the time decay of the net electron spin is monoexponential
\begin{equation}
\overline{S_z(t)} = S_z(0) \exp(- \overline{\gamma_{\tilde z}} t)
\end{equation}
and described by the domain-averaged rate
\begin{equation}\label{gamma_z_av}
\overline{\gamma_{\tilde{z}}} = \frac{\overline{\Omega_R^2(\bm{\rho})} \tau}{1+\sqrt{1+\Omega_D^2\tau^2}} \:.
\end{equation}

(iii) Small domains, $1/(\Omega_R^2 \tau) , 1/(\Omega_D^2 \tau) \gg \tau_c \gg \tau$. In this regime of spin dephasing, electrons move from one domain to another more frequently than the $\gamma_{\tilde{z}}$ and $\gamma_{\tilde{y}}$ rates. The in-plane component of the net electron spin does not emerge due to the frequent change in the Rashba field sign. The out-of-plane component exponentially decays at the rate 
\begin{equation}\label{Gamma_xx_av}
\overline{\Gamma}_{zz} = \overline{\Omega_R^2 (\bm{\rho})} \, \tau \: 
\end{equation}
which is independent of the Dresselhaus field strength. The dephasing rate~(\ref{Gamma_xx_av}) is twice as high as the rate~(\ref{gamma_z_av}) even at $\Omega_D\tau <1$.
 
(iv) $1/(\Omega_R^2 \tau) \gg \tau \gg \tau_c$. The correlation time of the Rashba field is shorter than the scattering time, and spin dephasing occurs at ballistic electron transport. Such regime is realized in high-mobility QW structures where the fluctuating Rashba field stems from the electric field of charged impurities in remote doping layers~\cite{Sherman03,Glazov10}. The spin dynamics of an electron with a certain wavevector $\bm{k}$ is described by  
\begin{equation}\label{dynr}
\frac{d\bm{s_k}}{dt} + \bm{s_k} \times [\bm\Omega_{D, \bm k} + \bm\Omega_{R, \bm k}(\bm{\rho}_0+\bm{v}t)] = 0 \:,
\end{equation}
where $\bm{\rho}_0$ in the initial electron coordinate, $\bm{v}=v_F (\bm{k}/k)$, and $v_F$ is the Fermi velocity. Equation~(\ref{dynr}) can be solved by
changing the spin coordinate frame to the one rotating along the $z$ axis with the angular frequency $\bm{\Omega}_{D,\bm{k}}$. The procedure is equivalent to the substitution $\bm{s_k} = \bm{R}_{\bm k} \, \tilde{\bm{s}}_{\bm k}$ and $\bm{\Omega}_{R, \bm k} = \bm{R}_{\bm k} \, \tilde{\bm{\Omega}}_{R, \bm k}$ with $\bm{R}_{\bm k}$ being the rotation matrix,
\[
\bm{R}_{\bm k} = \left[ 
\begin{array}{ccc} 
\cos[\Omega_D t (k_x/k_F)] & -\sin[\Omega_D t (k_x/k_F)] & 0 \\ 
\sin[\Omega_D t (k_x/k_F)] & \cos[\Omega_D t (k_x/k_F)] & 0 \\  
0 & 0 & 1
\end{array} 
\right] .
\]
This yields $d \tilde{\bm{s}}_{\bm k} / dt + \tilde{\bm{s}}_{\bm k} \times \tilde{\bm{\Omega}}_{R, \bm k} (\bm{\rho}_0+\bm{v}t)=0$. Taking into account that the Rashba field lies in the QW plane and $\Omega_{R} (\bm{\rho}) \tau_c \ll 1$  we obtain the dephasing rate of the $z$ spin component for electrons with the wavevector $\bm{k}$
\begin{equation}\label{eq:gr1}
\overline{\Gamma}_{zz}(\bm{k}) = \int\limits_0^{\infty} \overline{ \tilde{\bm\Omega}_{R, \bm k}(\bm{\rho}_0 + \bm{v}t) \cdot \tilde{\bm\Omega}_{R, \bm k}(\bm{\rho}_0)} \, dt  
\end{equation}
\vspace{-0.5cm}
\[
= \int\limits_0^{\infty} \overline{ \Omega_{R}(\bm{\rho}_0 + \bm{v}t) \, \Omega_{R}(\bm{\rho}_0)} \cos[\Omega_D t (k_x/k_F)] \,dt  \:,
\]
where the overline denotes averaging over the initial electron coordinate.

The spatial correlation function of the Rashba field caused by Coulomb impurities randomly distributed in the 
$\delta$-doping layers positioned at the distance $z_d$ below and above the QW has the form (see Ref.~\cite{Sherman03})
\begin{equation}\label{eq:omecor}
\overline{\Omega_R (\bm\rho) \, \Omega_R(\bm\rho')} = \frac{\overline{\Omega_R^2(\bm{\rho})}(2z_d)^3}{[(\bm\rho-\bm\rho')^2+(2z_d)^2]^{3/2}} \:.
\end{equation}
Combining Eqs.~(\ref{eq:gr1}) and~(\ref{eq:omecor}) one obtains
\begin{equation}
\overline{\Gamma}_{zz}(\bm{k}) = \overline{\Omega_R^2(\bm{\rho})} \, \Omega_{D} \tau_c^2 (k_x/k_F)  K_1[\Omega_{D} \tau_c (k_x/k_F)] \:,
\end{equation}
where $\tau_c$ is the correlation time of the Rashba field
\begin{equation}
\tau_c = \dfrac{\int\limits_0^{\infty} \overline{ \Omega_{R}(\bm{\rho}_0 + \bm{v}t) \, \Omega_{R}(\bm{\rho}_0)} \, dt}
{\overline{\Omega_{R}^2(\bm{\rho}_0)}} = \frac{2 z_d}{v_F} \:
\end{equation}
and $K_1$ is the modified Bessel function of the second kind. At $\Omega_D \tau_c \ll 1$, the rate $\overline{\Gamma}_{zz}(\bm{k})$ is given by $\overline{\Omega_R^2(\bm{\rho})} \tau_c$.

The dephasing rate of the net electron spin in the structure can be obtained by averaging $\overline{\Gamma}_{zz}(\bm{k})$
over the wavevector direction, which yields
\begin{equation}
\langle \overline{\Gamma}_{zz}(\bm{k}) \rangle = \overline{\Omega_R^2(\bm{\rho})} \tau_c \, \xi [ I_0(\xi)K_1(\xi) - I_1(\xi)K_0(\xi)] \:,
\end{equation}
where $I_{\nu}$ is the modified Bessel function of the first kind and $\xi=\Omega_D \tau_c /2$. The dephasing rate is given by $\overline{\Omega_R^2(\bm{\rho})} \tau_c$ (see Ref.~\cite{Sherman03}) and $\overline{\Omega_R^2(\bm{\rho})}/\Omega_D$ in the cases of moderate, $\Omega_D\tau_c \ll 1$, and strong, $\Omega_D\tau_c \gg 1$, Dresselhaus field, respectively.

\begin{figure}
\includegraphics[width=0.9\columnwidth]{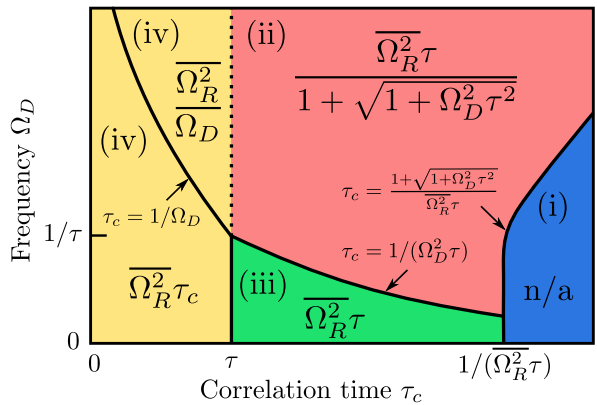}
\caption{Dephasing rate of the out-of-plane spin component in QWs with the fluctuating Rashba field in various regimes.}
\label{fig:summary}
\end{figure}
The theoretical results on spin dephasing rate in quantum wells with the fluctuating Rashba field are summarized in Fig.~\ref{fig:summary}. The strong Dresselhaus field increases the electron spin lifetime in all regimes of spin dymanics.

\section{Summary}

We have developed the microscopic theory of electron spin dephasing for asymmetric and macroscopically symmetric (110)-grown quantum wells where the Rashba field is homogeneous or fluctuates in the quantum well plane, respectively. Both low-mobility and high-mobility structures are analyzed. It is shown that the regime and rate of spin dephasing are determined by the strength of Rashba and Dresselahaus fields, the latter being always present in zinc-blende-type structures, as well as the electron mobility. The Dresselhaus field points along the growth direction in (110) quantum wells and increases the lifetime of the out-of-plane spin component. 

In asymmetric structures, the interplay of the homogeneous Rashba and Dresselhaus fields leads to a coupling of the in-plane and out-of-plane spin
components. The spin lifetime of the out-of-plane component nonmonotonically depends on the electron mobility. It linearly decreases with the scattering time $\tau$ in low-mobility structures, reaches the minimum at $\tau$ comparable to the period of spin precession in the effective magnetic field, and then linearly increases with $\tau$ in high-mobility structures.

In quantum wells with the fluctuating Rashba field, the scenario of spin dephasing is determined by the scale of the field fluctuations compared to other relevant scales. We show that, for all regimes of spin dynamics, the strong Dresselhaus field slows down the dephasing rate of the out-of-plane spin component.

The authors acknowledge fruitful discussions with M.\,M.\,Glazov. 
This work was supported by the RFBR, 
RF President Grants MD-2062.2012.2 and NSh-5442.2012.2, EU
projects POLAPHEN and SPANGL4Q, and the Foundation
``Dynasty''.


%

\end{document}